\documentstyle[hep93,epsfig]{article}
\def\refe#1{(\ref{#1})}
\def\bi{\begin{itemize}}
\def\ei{\end{itemize}}

\def\be{\begin{equation}}
\def\ee{\end{equation}}
\def\l{\label}
\def\ba{\begin{array}}
\def\ea{\end{array}}
\def\ss{\scriptscriptstyle}
\def\ltap{\ \raisebox{-.4ex}{\rlap{$\sim$}} \raisebox{.4ex}{$<$}\ }
\def\gtap{\ \raisebox{-.4ex}{\rlap{$\sim$}} \raisebox{.4ex}{$>$}\ }
\def\via{\vskip-\parskip \noindent}
\def\Rp{{R\!\!\!\!\!\:/}}
\def\GU{{\rm \scriptscriptstyle GU}}
\def\T{{\it \scriptscriptstyle tripl}}
\def\D{{\it \scriptscriptstyle doubl}}
\def\leff{\lambda_{333}'{}^{\!\!\!\!\!\! \rm \ss eff}}
\def\L{{\rm \scriptscriptstyle L \!\!\!\!\!\;/}}
\def\nuvev{\langle \tilde \nu_3 \rangle}

\def\npb#1#2#3{    {\it Nucl. Phys. }{\bf B #1} (19#2) #3}

\def\plb#1#2#3{    {\it Phys. Lett. }{\bf B #1} (19#2) #3}
\def\prd#1#2#3{    {\it Phys. Rev. }{\bf D #1} (19#2) #3}

\def\prl#1#2#3{    {\it Phys. Rev. Lett. }{\bf #1} (19#2) #3}

\def\mpla#1#2#3{   {\it Mod. Phys. Lett. }{\bf A #1} (19#2) #3}

\begin{document}
\title{$R$-Parity Breaking Phenomenology}
\author{Francesco Vissani \\
        {\em  International Centre for Theoretical Physics, ICTP }\\
        {\em  Strada Costiera 11, I-34100 Trieste, Italy;}\\
        {\em  Istituto Nazionale di Fisica Nucleare, INFN}\\
        {\em  Sezione di Trieste}
       }
\abstract{
We review various features of the 
$R$-parity breaking
phenomenology, with particular attention to the 
low energy observables, and to the 
patterns of the $R$-parity breaking interactions that 
arise in Grand Unified models.}
\maketitle

\section{Introduction}\label{sec:intro}
The supersymmetrization of the Standard Model (SM) requires
enlarging the spectrum of the theory. 
The quarks and leptons or the Higgs scalars become components
of supersymmetry group representations, the chiral superfields.
The notation is presented in table \ref{tab:notations}.
\begin{table}[htb]\label{tab:notations}
\begin{center}
\begin{tabular}{|c|c|c|c|} 
\hline
\footnotesize Chiral & \footnotesize Left Weyl & \footnotesize Complex 
  & $\scriptstyle SU(3)_c\times SU(2)_L\times U(1)_Y$ \\
\footnotesize Superfield  & \footnotesize fermion & \footnotesize scalar 
  & \footnotesize representation  \\ \hline
&&&\\
$L   $ &$ l         $ &$\tilde l$ &$(1,2,-1/2)$ \\
$E^c $ &$ e^c       $ &$\tilde e^c$ &$(1,1,1)$ \\
$Q   $ &$ q         $ &$\tilde q$ &$(3,1,1/6)$ \\
$U^c $ &$ u^c       $ &$\tilde u^c$ &$(3^*,1,-2/3)$ \\
$D^c $ &$ d^c       $ &$\tilde d^c$ &$(3^*,1,1/3)$ \\
$H_1 $ &$ \tilde h_1$ &$       h_1$ &$(1,2,-1/2)$ \\
$H_2 $ &$ \tilde h_2$ &$       h_2$ &$(1,2,1/2)$ \\
&&&\\
\hline
\end{tabular}
\end{center}
\caption{Chiral superfields and corresponding component fields.}
\end{table}
According to the usual convention we denote
the supersymmetric particles by a tilde.
Notice that supersymmetry implies that two Higgs 
doublets, $h_1$ and $h_2,$ are present.

Schemes of supersymmetry breaking allow us to generate 
superpartners mass patterns consistent with 
the present non-observation of superparticles. 
Theoretical arguments
require these masses to be below the TeV range,
making the supersymmetric extension of the
Standard Model interesting for present and future 
searches.

Each interaction of the Standard Model can be generalized 
in a supersymmetric invariant form. 
For instance the down Yukawa interactions read:
\be
{\cal L}_{\stackrel{\rm\ss Yukawa}{\rm\ss down}}=
-{ Y_D^*}_{ij} \ \left[\
h_1\ q_i\ d^c_j +\tilde h_1\ q_i\ \tilde d^c_j + \tilde h_1\ \tilde q_i\ d^c_j 
\right]\ + {\rm h.c.}
\l{eq:DownYukawa}
\ee
($Y_D$ are the down Yukawa couplings; 
$i,j$ are family indices;
$SU(2)_L$ doublets are contracted with $i\tau_2.$).
Due to supersymmetry, the SM interaction 
induces similar interactions between 
pairs of superpartners.

It is also easy to write interactions that
have no SM analogue.
This happens when superpartners behave as a dilepton,
$l_i l_j \tilde e^c_k;$ or as leptoquarks,
$l_i \tilde q_j d^c_k$ and $l_i q_j \tilde d^c_k;$
or finally as diquarks,
$\tilde d^c_i d^c_j u^c_k$ and $d^c_i d^c_j \tilde u^c_k$
(the supersymmetric form can be easily inferred in analogy  
with eq.\ \refe{eq:DownYukawa}).
We conclude that SM gauge invariance does not assure lepton and/or
baryon number conservation in supersymmetric context.
Notice that these interactions are not necessarily
linked to the supersymmetric breaking mechanism,
or to the structure of the Higgs sector,
about which we have not direct experimental informations yet,
but just depend on the spectrum of the model and on
SM gauge invariance. 

Some or all the interactions above
can be forbidden adding more symmetry to the model.
Such a symmetry can be local or global, continuos or discrete.
A widely used possibility is given by the $Z_2$ transformation 
upon which only the superpartner changes sign: the
$R$-parity. 
Since $R$-parity forbids the terms introduced above,
this assumption amounts to baryon {\em and} lepton 
number conservation\footnote{Even if
interactions which violate $R$-parity are forbidden, there is
the interesting possibility of spontaneous breaking of
the lepton number, due to the vacuum expectation value (VEV) of
the sneutrino \cite{spont}. 
Nonzero VEVs of other scalars would instead
break color and/or electric charge.}. 
One can even speculate about the origin of such a symmetry.
But there are still no experimental keys
to know which is the scheme chosen by Nature.
Therefore, a phenomenological attitude toward 
the supersymmetric paradigm requires to study
the consequences of relaxing the assumption of $R$-parity 
conservation.

The plan of the exposition is as follows:
First, we define the $R$-parity breaking 
interactions, and study their
possible manifestations and some experimental bounds.
We pay particular attention to the 
rare and exotic low-energy processes. 
Then, in an effort to obtain finer control 
on these interactions, we consider them in the 
context of Grand Unification (GU).
We discuss an interesting scenario, 
in which sizable $R$-parity 
breaking interactions can be reconciled 
with Grand Unification program.

\section{$R$-parity breaking}

In this section we define the $R$-parity breaking couplings,
and study possible manifestations of their presence.
We figure out important processes and give a feeling of the
existing bounds on the couplings. The possibility-risk 
that $R$-parity breaking couplings make the ordinary matter 
unstable is analyzed.

\subsection{Definitions and fundamental facts.}

The superpotential $W$
gives us a compact formalism to describe the 
supersymmetric interactions of matter fields:
By definition it is an analytic function 
of the chiral left superfields present in the theory.
Let us decompose $W=W_R+W_\Rp.$ 
Considering the renormalizable
interactions, the $R$-parity conserving part reads:
\be
W_R=\frac{m_{e_i}}{v_1} H_1 L_i E^c_i +
    \frac{m_{d_i}}{v_1} H_1 Q_j V_{ji}^* D^c_i -
    \frac{m_{u_i}}{v_2} H_2 Q_i U^c_i +
    \mu H_1 H_2
\l{eq:R-parity-conserving}
\ee
whereas the $R$-parity violating part reads:
\be
W_\Rp= \lambda_{ijk} L_i L_j E^c_k
      +\lambda_{ijk}' L_i Q_j D^c_k
      +\lambda_{ijk}'' D^c_i D^c_j U^c_k
      +\mu_i L_i H_2
\l{eq:R-parity-violating}
\ee
The superpotential $W$
is written in terms of superfields with
fermion mass eigenstates, so that the Cabibbo-Kobayashi-Maskawa
matrix $V_{ij}$ appears in \refe{eq:R-parity-conserving} and
\refe{eq:R-parity-violating} explicitly as well as 
in $Q_i=(U_i, V_{ij} D_j)$;
$m_{e_i},$ $m_{d_i},$ $m_{u_i}$
are the fermion masses.
Finally, $v_{1,2}$ are the vacuum
expectation values of the scalar components of 
the superfields $H^0_{1,2}.$

Notice that with a proper redefinition,
$\mu H_1 + \mu_i L_i \to \mu H_1$
the last term can be eliminated from the 
superpotential. Therefore we will assume in the following $\mu_i=0.$
In passing, we remark that the presence of all but the third term
in \refe{eq:R-parity-violating} is due to the fact that the 
Higgs $H_1$ and the three lepton doublets $L_i$ are identical from the point
of view of gauge symmetry.

$W_R$ is by definition 
the superpotential of the Minimal Supersymmetric
extension of the SM (MSSM). 
Notice that in full analogy with the SM case \refe{eq:R-parity-conserving}
conserves the four $U(1)$ numbers related to $B$ (baryon charge),
${L_1},$ ${L_2}$ and ${L_3}$ (lepton charges)
where the definitions are 
done on the superfields\footnote{This definition is forced by the 
gaugino interactions. 
Notice that the scalar masses can  
provide us with sources of violation of hadronic and leptonic flavours, 
if they are not diagonal in the same basis in 
which the fermion masses are diagonal.}.
Let us therefore analyze 
the interactions in \refe{eq:R-parity-violating}
from the point of view of the global symmetries.
They violate either total lepton ($\lambda,\lambda',\mu_i$) or
baryon number symmetries ($\lambda''$). 
One can further divide in two classes the lepton violating terms:
The terms in the first class,
\be
L_i H_2,\ \ \ L_i Q_k D^c_l,\ \ \ L_i L_j E^c_j,\ \ \ \ \ \ \ i \neq j   
\l{eq:class1}
\ee
carry charges ${L_i}$, whereas those in the second class,
\be
L_1 L_2 E^c_3,\ \ \ L_3 L_1 E^c_2,\ \ \ L_2 L_3 E^c_1
\l{eq:class2}
\ee
carry charges ${\cal L}_3=L_1+L_2-L_3,$ 
${\cal L}_2=L_1-L_2+L_3$  and
${\cal L}_1=-L_1+L_2+L_3$ respectively.
This classification has some importance for lepton violating
phenomena.
For instance, the neutrino mixing term $\nu_1\nu_2$ 
cannot be generated by the operators \refe{eq:class2} alone,
since its charge is $1/2\ ({\cal L}_1+{\cal L}_2) + {\cal L}_3$ 
(it would requires ``half" vertices; similarly for 
the other mixings). 
For the same reason the terms \refe{eq:class1} 
cannot be induced by those of \refe{eq:class2} alone.

Few remarks, in order to give a perspective to the present study.\via
(1) It is of course possible to ascribe $B$- and $L$-violating phenomena
to $R$-parity conserving theories, for example in the case of 
supersymmetric $SU(5)$ model; but, 
due to the different underlying mechanism,
the resulting phenomenology 
is typically different.\via 
(2) One may consider the situation in which $R$-parity is a 
symmetry of the tree level lagrangian, broken by effective terms. 
It has been remarked \cite{Hallreview} that if we want to reconcile 
the theory with a global invariance, we have to consider 
operators at least of dimension 7, for example $(L H_2)^3$ (which is 
invariant under a leptonic $Z_3$). Previous argument 
does not however disfavor this scenario, 
since what matters is the symmetry
of the underlying fundamental theory \cite{K}.\via
(3) There is the possibility that supersymmetric
interactions are $R$-parity symmetric, 
whereas the interactions which break supersymmetry are not. 
To our knowledge, this possibility has not attracted 
a lot of attention. A case of special interest is 
studied in \cite{IHLee}.

\subsection{Exotic interactions of ordinary matter}

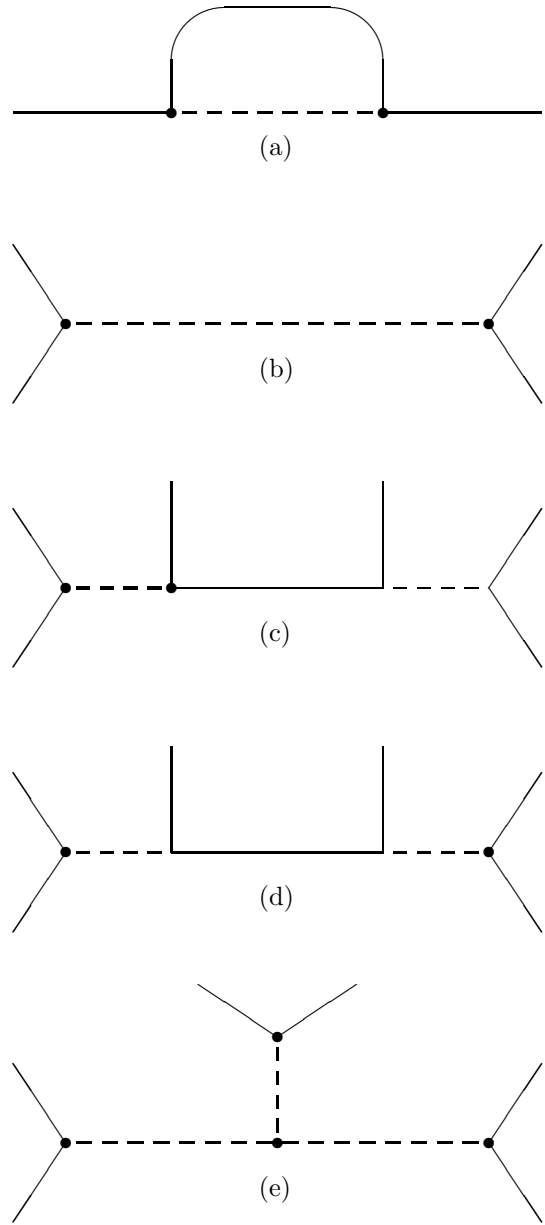
\begin{figure}[h]

\begin{center}

\begin{picture}(200,500)(0,0)

\put(93,414){(a)}
\put(93,330){(b)}
\put(93,230){(c)}
\put(93,130){(d)}
\put(93,20){(e)}

\put(0,430){\line(1,0){60}}
\multiput(64,430)(9.5,0){8}{\thicklines \line(1,0){5.5}}
\put(100,430){\oval(80,80)[t]}
\put(200,430){\line(-1,0){60}}

\put(60,430){\circle*{4}}
\put(140,430){\circle*{4}}

\put(20,350){\line(-2,-3){20}}
\put(20,350){\line(-2,3){20}}
\multiput(24,350)(9.75,0){16}{\thicklines \line(1,0){5.75}}
\put(180,350){\line(2,-3){20}}
\put(180,350){\line(2,3){20}}

\put(20,350){\circle*{4}}
\put(180,350){\circle*{4}}

\put(20,250){\line(-2,-3){20}}
\put(20,250){\line(-2,3){20}}
\multiput(24,250)(9,0){4}{\thicklines \line(1,0){5}}
\put(60,250){\line(1,0){80}}
\multiput(144,250)(9,0){4}{\line(1,0){5}}
\put(180,250){\line(2,-3){20}}
\put(60,250){\line(0,1){40}}
\put(140,250){\line(0,1){40}}
\put(180,250){\line(2,3){20}}

\put(20,250){\circle*{4}}
\put(60,250){\circle*{4}}

\put(20,150){\line(-2,-3){20}}
\put(20,150){\line(-2,3){20}}
\multiput(24,150)(9,0){4}{\thicklines \line(1,0){5}}
\put(60,150){\thicklines \line(1,0){80}}
\multiput(144,150)(9,0){4}{\thicklines \line(1,0){5}}
\put(180,150){\line(2,-3){20}}
\put(60,150){\line(0,1){40}}
\put(140,150){\line(0,1){40}}
\put(180,150){\line(2,3){20}}

\put(20,150){\circle*{4}}
\put(180,150){\circle*{4}}

\put(20,40){\line(-2,-3){20}}
\put(20,40){\line(-2,3){20}}
\multiput(24,40)(9.75,0){16}{\thicklines \line(1,0){5.75}}

\put(100,80){\line(-3,2){30}}
\put(100,80){\line(3,2){30}}
\multiput(100,44)(0,9){4}{\thicklines \line(0,1){5}}

\put(180,40){\line(2,-3){20}}
\put(180,40){\line(2,3){20}}

\put(20,40){\circle*{4}}
\put(180,40){\circle*{4}}

\put(100,40){\circle*{4}}
\put(100,80){\circle*{4}}

\end{picture}

\end{center}
\caption{Topologies of Feynman diagrams 
involving $R$-parity
breaking couplings (represented by the blobs)
which induce important interactions among the 
Standard Model particles. Fermions (bosons) are indicated
by continuos (dashed) lines.
\label{fig:topologies}}
\end{figure}

Let us consider the effective
terms that SM inherits from the $R$-parity breaking interactions
when the sleptons 
and the squarks fields are integrated away\footnote{There is
an important consequence of $R$-parity breaking interactions
regarding the supersymmetric particles:
the lightest supersymmetric particle becomes unstable.
See \cite{Dreiner} for searches at colliders.}.
The topologies of Feynman diagrams that is necessary 
to consider are listed in figure \refe{fig:topologies}.

The operators of greatest interest are clearly those which violate
SM conservation laws, the lepton and/or the baryon numbers, or 
give flavor-changing neutral currents.
In the case of the two fermion operators there are the
Majorana neutrino masses $\nu\nu$ (fig.\ \ref{fig:topologies}a);
for the six fermions operators, either those of the form
$e u \bar d e u \bar d$ 
which trigger neutrinoless double beta decay (fig. \ref{fig:topologies}c,
\ref{fig:topologies}d),
or those of the form 
$u d d u d d,$ which give for instance $n$-$\bar n$ oscillations
(fig.\ \ref{fig:topologies}c,\ref{fig:topologies}e)
\footnote{For further informations see references
\cite{neutrino-masses}, \cite{0nu-bb} and \cite{n-nbar}.}.
We recall that the first two types of operators
arise in pure lepton number violating framework, 
whereas the last just requires violation of baryon number;
notice also that their flavor 
structure can be {\em a priori} generic.

Now let us focus the attention on 
the four fermions operators, arising 
by diagrams of the topology of
fig.\ \ref{fig:topologies}b. 
They are listed in table 2,
together with the couplings involved, the particle exchanged
and a typical process triggered. 
\begin{table}[htb]
\begin{center}
\begin{tabular}{|c|c|c|c|} 
\hline
\footnotesize Effective
& \footnotesize Particle
& \footnotesize Couplings 
& \footnotesize Example \\
\footnotesize operator 
& \footnotesize exchanged
& \footnotesize involved 
& \footnotesize process \\
\hline
&&&\\
$e e \bar e\bar e   $ 
& $\tilde \nu  $ 
& $\lambda^2 $ 
&  $\mu^-  \to e^- e^- e^+  $ \\
$e \nu \bar e \nu   $ 
& $\tilde e,\tilde e^c,\widetilde{ ee^c  }$ 
& $\lambda^2 $ 
&  $ \mu^- \to e^- \nu_e \bar \nu_\mu $ \\
$d d \bar d \bar d   $ 
& $\tilde \nu,\tilde u^c $ 
&${\lambda'}^2, {\lambda''}^2$ 
& $  K^0-\bar K^0   $ oscill. \\
$u d \bar u \bar d   $ 
& $\tilde e,\tilde d^c $ 
&${\lambda'}^2, {\lambda''}^2$ 
& $ B \to  $ non charmed \\
$u e \bar u \bar e   $ 
& $\tilde  d^c $ 
&${\lambda'}^2$ 
& $ D^+ \to \pi^+ \mu e  $ \\
$d \nu \bar d \nu   $ 
& $\tilde d,\tilde d^c,\widetilde{ dd^c  }$ 
& ${\lambda'}^2$ 
&  $ B \to K \nu \bar \nu $ \\
$u e \bar d \nu   $ 
& $\tilde e,\widetilde{ ee^c} ,\tilde d,\widetilde{ dd^c } $ 
& ${\lambda'}^2,\lambda{\lambda'}^2$ 
&  $ B \to K l \bar \nu  $ \\
$d e \bar d \bar e   $ 
& $\tilde\nu,\tilde u  $ 
& ${\lambda'}^2,\lambda{\lambda'}^2$ 
& $ K_L \to \mu e $ \\
$u u d e   $ 
& $\tilde d^c $ 
& $\lambda' \lambda''$ 
& $ p \to \pi^0 e^+ $ \\
$u d d \nu   $ 
& $\tilde d^c,\widetilde{ dd^c }$ 
& $\lambda' \lambda'' $ 
& $ p \to K^+\nu $ \\
$d d d \bar e   $ 
& $\widetilde{ uu^c }$ 
& $\lambda' \lambda''$ 
& $ n \to K^+ e^- $ \\
&&&\\
\hline
\end{tabular}
\end{center}
\caption{Four fermions operators resulting from $R$-parity
breaking interactions. 
In first column 
$\nu$ denotes either the neutrino {\em or} the 
antineutrino field.
The propagators like $\widetilde{ ee^c}$ in second column 
arise from the mixing of the scalar states $\tilde e$ 
and $\tilde e^c$ after $SU(2)_L$ breaking.}
\end{table}

The most important operators are clearly those of last three
rows of table 2,
since they lead to instability of nucleons. 
As it is well known, they arise due to violations of {\em both} the 
baryon and the lepton number.
The fact that there are no four-fermion operators 
which violates only the baryon number
is a general consequence of
$SU(3)_c\times U(1)_{e.m.}$ symmetry.
Violations of the 
lepton number are possible,
but only in the interactions involving neutrinos:
As an example, the exchange of $\widetilde {\tau\tau^c}$
induces the decay $\mu^-\to e^- \nu_e \bar \nu_\mu$
due to $\lambda_{123}$ and $\lambda_{231}$ couplings\footnote{
Unfortunately, existing limits on the single couplings 
render this process not experimentally interesting in the 
model under consideration---I thank M.\ Cooper for
a clarification about this point.}.

Table 2 illustrates the need to proceed carefully
in introducing the $R$-parity violating couplings,
since all kind of non-standard operators can be  induced.
According to previous observation, a safe possibility 
of introducing the $R$-parity violating terms
is to forbid the $B$-violating terms, but to retain the lepton 
violating ones (or viceversa). 
A more daring possibility is to have both operators to a sufficiently 
suppressed level.
We discuss these possibilities in the following, 
with particular attention 
to the manifestations in the low-energy physics.

\subsection{Lepton-violating scenario}\label{sub:lviol}

Suppose for a while that $B$-violating terms are absent.
Consider the flavor structure of the 
$R$-parity breaking couplings.
Are there couplings unconstrained by rare (or forbidden)
processes? 
A partial answer is provided by table 3.
\begin{table}[htb]
\begin{center}
\begin{tabular}{|c||c|c|c||} \cline{2-4}
\multicolumn{1}{c||}{ }   
& $K^0-\bar K^0$ & $B^0-\bar B^0$ & $K_L\to \mu e$  \\ 
\hline \hline 
111 & & & x    \\ \hline 
112 & x & & x  \\ \hline 
121 & x & & x  \\ \hline 
211 & & & x    \\ \hline 
122 & & & x    \\ \hline 
212 & x & & x  \\ \hline 
221 & x & & x  \\ \hline 
222 & & & x    \\ \hline 
113 & & x &    \\ \hline 
131 & & x & x  \\ \hline 
311 & & &      \\ \hline 
123 & & x &    \\ \hline 
132 & & x & x  \\ \hline 
213 & & x &    \\ \hline 
231 & & x & x  \\ \hline 
312 & x & & x  \\ \hline 
321 & x & & x  \\ \hline 
223 & & x &    \\ \hline 
232 & & x & x  \\ \hline 
322 & & &      \\ \hline 
133 & & &      \\ \hline 
313 & & x &    \\ \hline 
331 & & x &    \\ \hline 
233 & & &      \\ \hline 
323 & & x &    \\ \hline 
332 & & x &    \\ \hline 
333 & & &      \\ \hline 
\hline
\end{tabular}
\end{center}
\caption{Rare processes in which the various $\lambda_{ijk}'$
couplings are involved. 
A sneutrino or an up squark is exchanged.}
\end{table}
It shows whether three ``delicate'' observables 
can be affected or not by
the $\lambda'$-type couplings 
(the precise meaning of the table is:
whether the coupling enters or not a tree level diagram 
relevant for the processes).

We deduce from table 3 that the couplings 
$\lambda'_{3jj}$ and $\lambda'_{j33}$ 
do not give contribution to the processes. 
This means that large values of these couplings are
not incompatible with present experimental informations. 
As  a common feature, these couplings do not violate 
hadronic flavours.  

We can somewhat push the above argument. 
Let us suppose that one $\lambda'$ coupling, 
which is not in the class above, is large.
Table 3 tell us that, in this case,
some other $R$-parity couplings is
constrained by present experimental bounds.
To be quantitative, the observation of whichever coupling
$\lambda'_{\rm obs}$ at the level of $10^{-2}$ would imply a strong
suppression ($\gtap 10^{-4} \times \lambda'_{\rm obs}$) 
of another $\lambda'$ coupling.
In absence of a theoretical explanation, this scenario is
questionable on the basis of naturalness.

\subsection{Lepton- and baryon-violating  scenario}\label{sub:lbviol} 

The simultaneous presence of the couplings 
$\lambda''$ ($B$-violating) and $\lambda'$ ($L$-violating)
leads to the possibility of squark-mediated
proton decay.
This implies very strong bounds on the couplings which allow 
the decay at tree level:
\be
|\lambda'\cdot \lambda''| \ltap 10^{-24}
\l{proton-decay-bound-on-lambda}
\ee
for squark masses around 1 TeV \cite{HK}.

The bound does not affect certain couplings 
involving heavy generations. 
But, since the bounds are so stringent, it is
important to check the one loop structure of the theory.  
It is possible to prove that, choosing
whichever pair of couplings $\lambda'$ and $\lambda'',$ 
there is always at least one diagram relevant for
the decay at one loop level \cite{Upper}. 
This happens due to the flavor-changing interactions, 
which are present even in the absence of $R$-parity breaking,
namely: the interactions of the quarks with 
the $W$ boson and the charged Higgs, and their supersymmetric
counterparts. The less suppressed pair of couplings 
is still subject to a (conservative) bound on their product,
\be
|\lambda'\cdot \lambda''| \ltap 10^{-9},
\ee
according to \cite{Upper}\footnote{A different
conclusion has been reached by \cite{CRS}.}. 
The simultaneous presence of suitably chosen
couplings $\lambda$ and $\lambda''$
seems instead to be less dangerous for 
proton decay \cite{CRS}. 
Notice for instance that, due
to the symmetry discussed above, 
the presence of operators of the class 
\refe{eq:class2} requires that there are three different leptons 
in the final states, calling for dimension 9 
effective operators for nucleon decay.

Coming to phenomenology, we remark that the squark-mediated 
nucleon decay may have a very neat experimental signature: 
the  presence of the $(B+L)$-conserving 
channels\footnote{Another possible manifestation of
these dynamics would be the presence of  unexpected branching-ratios
for the $(B-L)$-conserving channels of nucleon decay.} 
\cite{BplusL}.
These channels are related to effective operators at least of 
dimension 7 \cite{dim7}.
This calls for sources of $SU(2)_L$ breaking, 
which are provided by left-right squark mass mixing: for the 
top quark we have
$m_{\tilde t \tilde t^c}^2 \sim m_t\ \tilde m$
where $m_t$ is the top mass, that is not expected to be 
very different from the typical supersymmetric mass
$\tilde{m}.$
Regardless of the Lorentz structure,
there is only one effective four-field operator at the quark level
which mediate $(B+L)$ conserving nucleon decays:
$d d s \bar l,$ where $l=e,\mu.$ 
It gives rise to $n \to K^+ l^-$ and 
$p \to K^+ l^- \pi^+$ decay channels. 
The first decay, which proceeds with a faster rate,
provokes the decay of the neutrons in the stable nuclei.
This  provide us with a quite clear signal
in water \v{C}erenkov detectors:
\begin{equation}
{}^{16}O\to {}^{15}O + \gamma(6.2\ {\rm MeV}) + \mu + l,
\label{exp-sign}
\end{equation}
where $l$ is monochromatic, $\mu$ results from kaon decay
and $\gamma$ from the transition of the excited nucleus
to the ground state (a unobservable neutrino from $K$ decay 
is also present).

A final remark.
Even if it is allowed to speculate on the possibility of very small
couplings, it would be much nicer to have 
a theoretical guideline to explain the size of the couplings.
In the context of horizontal symmetry \cite{HK,HOR}, the smallness
of the couplings can be related to suitably large horizontal charges. 
In our opinion however a defect of these approaches is that they 
still suffer of considerable latitude in the specification 
of the models.

\section{Supersymmetric Grand Unification and $R$-parity breaking}
\label{sec:susygut}

In previous sections we assumed that:\via
$(i)$ The Standard Model must be embedded into a supersymmetric theory;\via
$(ii)$ all the interactions compatible with the 
gauge symmetry should be a priori present. \via
Unfortunately, at present, hypothesis $(i)$ lacks 
of experimental support.
This requires to convey special attentions to the 
theoretical motivations for supersymmetry.
Among them, it is prominent the possibility to 
implement in the supersymmetric context
the Grand Unification program (in its minimal form).
Therefore we will further specify the theoretical context, 
and assume that: \via
$(iii)$ The interactions of the supersymmetric Standard Model 
are the low energy manifestations of a $SU(5)$ invariant dynamics.\via
This hypothesis  of course implies a specification of
the $R$-parity breaking couplings.

In the $SU(5)$ model one can introduce the following
$R$-parity violating interactions \cite{Ramond}
\be
\Lambda_{ijk} \bar 5_i \bar 5_j 10_{k} +
\bar 5_i(M_i + h_i \Phi) H,
\l{eq:lambda-su5}
\l{eq:R-viol-su5}
\ee
where $i,j,k=1,2,3$ are generation indices,
$\Lambda_{ijk}$ are the coupling constants and
$\bar 5_i,$ $10_i$ are the matter superfields which
can be written (restating the gauge indices) as:
\be
\bar 5^a = \left(
               \begin{array}{c}
                   D^{c\alpha}\\
                   \epsilon^{AB} L_B
               \end{array}
           \right)
\ \ \ \ \ \ \ \ \ \
10_{ab} = \left(
               \begin{array}{cc}
               \epsilon_{\alpha\beta\gamma}U^{c\gamma}& -Q_{B\alpha}\\
               Q_{A\beta}& \epsilon_{AB} E^c 
               \end{array}
      \right).
\l{eq:schematic-notation}
\ee
where $\epsilon^{12}=\epsilon_{21}=1.$
$M_i$ are mass parameters, $h_i$ are couplings,
$\Phi$ and $H$ are the 24-plet and 5-plet
of Higgs multiplets.
Starting from \refe{eq:R-viol-su5}, we will study in the 
following two possible scenarios for the 
$R$-parity breaking couplings.

\subsection{A model with small $R$-parity breaking 
couplings}\label{sub:lambdamodel}

We first consider 
the effects of $\Lambda$ couplings, in a model in which
the matter-Higgs mixing (the second term in \refe{eq:R-viol-su5}) is
negligible.

It is convenient to define
$\Lambda_{ijk}$ in the basis
where $SU(2)_L$-singlets $u^c$ and $d^c$
coincide with mass eigenstates. This always can be done since
$u^c$ and $d^c$ enter different $SU(5)$-multiplets.
Note that due to the antisymmetry of
10-plets the interactions \refe{eq:R-viol-su5} are antisymmetric
in generation indices: $\Lambda_{ijk}=-\Lambda_{jik}.$

Substituting the multiplets \refe{eq:schematic-notation} in
\refe{eq:R-viol-su5}
and performing the redefinitions of the couplings
which bring the $R$-parity conserving part of the superpotential 
with light fields in the form \refe{eq:R-parity-conserving}, 
we find the relations between original $\lambda_{ijk}$ and 
$\Lambda_{ijk}$ couplings
at the GU scale: 
\be
\begin{array}{l}
\lambda_{ijk}=-\Lambda_{i'j'k'}\ {\cal U}_{i'i}\ {\cal U}_{j'j}\ 
{\cal V}_{k'k}\\
\lambda'_{jki}=2\Lambda_{ij'k'}\ {\cal U}_{j'j}\ {\cal W}_{k'k} \\
\lambda''_{ijk}=\Lambda_{ijk}.
\l{eq:lambda-unified}
\end{array}
\ee
where ${\cal U},{\cal W},{\cal V}$ are unitary matrices. 
The appearance of these matrices can be explained 
considering that our choice of flavor basis does not fix the 
flavor structure of the superfield $L$ (respectively $E^c$ and $Q$) 
which appears together with $D^c$ ($U^c$) in the $SU(5)$
$\bar 5$-plet (10-plet).
They can be calculated fixing the mechanism of mass generation: 
which Higgs representation are present,
which non-renormalizable operators, {\em etc.}. 
We will consider the case:
\be
\begin{array}{c}
{\cal U}={\cal W}=1,\\ 
{\cal V}=V 
\end{array}
\l{eq:minimal}
\ee
which corresponds to the assumption that only 
Higgs 5-plets contribute to the fermion mass matrices.

As a consequence of
quark and lepton unification in $SU(5),$ all types of $R$-parity
violating couplings
appear simultaneously. Moreover, different couplings
$\lambda,\lambda'$ and $\lambda''$ are determined by unique
GU coupling $\Lambda.$ As follows from
\refe{eq:lambda-unified} and \refe{eq:minimal}, these couplings 
basically coincide at GU scale:
\begin{equation}
-\lambda_{ijl} V^{-1}_{lk}=\frac{1}{2} \lambda'_{jki}=\lambda''_{ijk}.
\l{eq:lambda-unified2}
\end{equation}
Notice that Grand Unification implies  that the $L$-violating
couplings $\lambda'_{ijk}$
should be antisymmetric in the exchange of the first and third indices:
$\lambda'_{ijk}=-\lambda'_{kji},$ similarly
to other couplings; 
in the non-unified version
\refe{eq:R-parity-violating} these couplings can have also a symmetric part.

The considerations above apply to the low energy theory
up to minor modifications.
A not completely negligible effect is the 
evolution of the couplings due to gauge renormalization.
It leads to modification of GU relations \refe{eq:lambda-unified2}
at the electroweak scale:
\be
\begin{array}{l}
\lambda_{ijk}=-1.5\ \Lambda_{ijl} V_{lk}\\
\lambda'_{jki}=2\ (3.4 \pm 0.3)\ \Lambda_{ijk}\\
\lambda''_{ijk}=(4.4 \pm 0.4)\ \Lambda_{ijk} 
\l{eq:lambda-running}
\end{array}
\ee
(the errors correspond to the uncertainty in strong coupling
constant:\ $\alpha_s(M_Z)=0.12\pm 0.01$).
The inclusion of other
uncertainties related {\it e.g.}\ to threshold
SUSY and GU corrections may require the doubling of the errors quoted.
The renormalization effects due to third family
Yukawa couplings do not drastically change the relations
\refe{eq:lambda-running}.

With previous remarks in mind, 
it is easy to understand that the couplings are subject to 
quite strong constraints from the proton decay bounds in the
case under consideration.
To be concrete, let us consider the bound on the coupling $\Lambda_{233}$  
(which may be argued to be the dominating one). 
The proton decay, induced at the one loop level, implies \cite{R-GUT}:
\be
\Lambda \ltap 3\cdot 10^{-9}
\ee
This can be thought as a conservative bound in this kind of GU models
for the $R$-parity breaking couplings. 
We conclude that, whereas present model easily encompasses
nucleon instability phenomena (in particular decays which conserve
$B+L,$ or decays with exotic branching ratios), 
it cannot account for large $R$-parity breaking couplings.

\subsection{A model with large $R$-parity 
breaking couplings}\label{sub:d-tmodel}

Let us consider a model where the
matter-Higgs mixing is  the only source of
$R$-parity violation.  
Suggesting that 
third generation coupling dominates,  we can
write the appropriate terms of
the superpotential in the following way
\be
\bar 5_3 \hat m H +
\bar H \hat M H + y_{i}\ \bar 5_i 10_i \bar H ,
\l{sup}
\ee
where $\bar 5_i$ and $10_i$ are defined in the diagonal basis for down quark
Yukawa couplings $y_i$, $i=d,s,b$
so that $d^c_i$ and $d_i$ coincide,
up to corrections $M_W/M_\GU,$
with mass eigenstates. 
The mass matrices  of \refe{sup}
can be written in the doublet-triplet form as:
\be
\hat m = {\rm diag}(m_\T, m_\D), \ \ \ \hat M = {\rm diag}(M_\T, M_\D),
\l{matr}
\ee
where $M_\T \sim M_\GU$ and $M_\D$, $m_\D$ and $m_\T$
are at the electroweak
scale (large value of  $m_\T$ would result in the fast proton decay).
The explanation of this mass pattern is clearly connected
to the explanation of the doublet-triplet (DT) 
splitting\footnote{We will 
not specify any underlying mechanism for DT splitting, 
but simply observe that it is technically possible to implement it
in the present context,  
carefully choosing $M_i,$ $h_i$ and $\langle \Phi\rangle $ 
in \refe{eq:R-viol-su5}.}. 

The first term in \refe{sup} can be eliminated by
rotations of the doublet and the triplet components of the
5-plets: $\bar 5_3=(B^c,L_3)$ and $\bar H=(\bar {\cal T}, H_1)$.
For triplet components we redefine:
\be
\begin{array}{rcl}
c_\T \bar {\cal T} + s_\T B^c & \to & \bar {\cal T} \\
c_\T B^c - s_\T \bar {\cal T} & \to & {B^c} ,  
\end{array}
\l{trip}
\ee
so that ${B^c}$ and $\bar {\cal T}$ are the mass states,
$c_\T \equiv \cos \theta_\T$,
$s_\T \equiv \sin \theta_\T,$  and
\be
\frac{s_\T}{c_\T} = \frac {m_\T}{M_\T}.
\l{trip-mix}
\ee
For doublet components:
\be
\begin{array}{rcl}
c_\D H_1 + s_\D L_3 & \to & H_1 \\
c_\D L_3 - s_\D H_1 & \to & L_3
\end{array}
\l{doub}
\ee
and
\be
\frac{s_\D}{c_\D} = \frac {m_\D}{M_\D}.
\l{doub-mix}
\ee
Since $m_\D, m_\T, M_\D\sim M_W$ one gets
from \refe{doub-mix}
and \refe{trip-mix} that
$s_\T$
is strongly suppressed, $s_\T \sim M_W/ M_\GU < 10^{-14}$, 
whereas $s_\D$ can be of the order 1.

Substituting the expressions \refe{trip} and \refe{doub} into 
\refe{sup} we obtain the effective $R$-parity violating
couplings \refe{eq:R-parity-violating}.
In particular the third generation Yukawa coupling gives
\be
\leff L_3 B^c Q_3',
\l{lambda333}
\ee
where
\be
\leff = s_\D\cdot y_b,
\ee
and $Q_3' \equiv V_{ib}^* Q_i$. Baryon violating interactions as well
as pure leptonic terms are absent due to the antisymmetry.
The Yukawa coupling of the  second generation leads to
\be
y_s\ 
[ s_\T  B^c S^c U^c_i +
  s_\D  L_3 S^c Q_i   +
  s_\D  L_2 L_3 E^c_i
]
\l{sec}
\ee
(The first generation Yukawa coupling gives similar terms with
the substitution $y_s V_{is}\to y_d V_{id},$
$S \to D$, $L_2 \to L_1$).

The leading contribution to the proton decay is induced by
$L$-violating interaction
\refe{lambda333} and $B$-violating interaction \refe{sec}. The
$\tilde{b}^c$ exchange dressed by  $h^+$, $\tilde{h}^+$...
results in
the amplitude for proton decay
\be
A \propto
\leff \cdot y_s\ s_\T\cdot \xi
= y_s y_b\ s_\D s_\T\ \xi ,
\ee
where $\xi$ is the loop suppression factor.
Substituting values of parameters,  
we find that even for large $\tan\beta$
($y_b \sim 1$) this amplitude is small enough to allow for
$s_\D$, and consequently,  $\leff$
to be of the order 1. All other diagrams give
smaller contributions. (Note that in the considered example all
the $B$-violating interactions contain $b^c$ quark, so that
even lowest family couplings need a loop ``dressing").

\subsection{Neutrino masses and large 
$R$-parity breaking couplings}\label{sub:neud-t}

There is another consequence of the 
matter-Higgs mixing \cite{R-GUT,banks,nir,H}:
explicit $R$-parity violating terms in
\refe{sup} induces in general VEV 
of sneutrino.
Indeed, the relevant terms in the potential at the electroweak
scale are:
\be
\ba{rl}
V \ni & (m_{L_3}^2+\delta m^2)\ |h_1|^2 + 
m_{L_3}^2\ |\tilde l_3|^2\ - \\[1ex] \nonumber
&[B\cdot M_\D\ h_1 h_2 + (B+\delta B)\cdot m_\D\ 
\tilde l_3 h_2 + {\rm h.c.}].
\l{scalar-potential}
\ea
\ee
To proceed in the discussion, we assume a definite 
scenario for supersymmetry breaking: the low-energy 
supergravity model.
We suggest that soft breaking terms are universal
at the scale $M_\GU$ suggested 
by gauge coupling unification.
Then the parameters $\delta m^2$ and $\delta B$ 
\refe{scalar-potential}
describe the renormalization effect due to the bottom
Yukawa coupling from $M_X$ to the electroweak scale.
The corresponding renormalization group equations are:
\be
\ba{cl}\displaystyle
\frac{d}{dt} \delta B =&  
3\ y_b^2\ A_b, 
\\[1.5ex] \nonumber \displaystyle
\frac{d}{dt} \delta m^2 =& 
3\ y_b^2\ (m_{Q_3}^2+m_{D^c_3}^2+m_{H_1}^2+A_b^2), 
\l{ren-group-for-breakers}
\ea
\ee 
where $t=1/(4\pi)^2 \times \log(M_\GU^2/Q^2).$
The rotation \refe{doub} 
which eliminates matter-Higgs mixing 
term in the superpotential generates mixing terms
for the sleptons:
\be
V_\L \approx  \theta_\D \times 
\left[ 
\delta m^2\ h_1^* + \delta B\cdot \mu\ h_2 
\right] \tilde l_3 + 
{\rm h.c.}
\l{lepton-violating-part}
\ee
(for small $\theta_\D$).
After electroweak symmetry breaking 
these mixing terms, together with soft 
symmetry breaking masses, 
induce a VEV of tau sneutrino of the order:
\be
\nuvev
\sim v\ \theta_\D\times  
\left(\frac{\delta m^2}{m_{L_3}^2}\ \cos\beta +
\frac{\delta B\cdot \mu}{\; m_{L_3}^2} \sin\beta \right).
\l{tau-sneutrino-vev}
\ee 
The factor in brackets can be estimated as
$y_b^2\ (3\ \cos\beta + 0.5\ \mu/m_{L_3}\;\sin\beta),$ 
where the figures quoted 
arise from approximate integration of renormalization group equations
\refe{ren-group-for-breakers}. Consequently
the tau sneutrino  
VEV is\footnote{ 
Technically it is possible to implement
a cancellation between the two terms in \refe{tau-sneutrino-vev} 
(see \cite{IHLee} for a phenomenological
study of such a possibility). However 
we see no natural reason for this to 
happen in the supergravity context.} 
$\nuvev\sim v\ \theta_\D\ y_b^2$. 
Due to this VEV the tau neutrino mixes with the zino,
and consequently the mass of tau neutrino is generated 
via the see-saw mechanism:
\be
\frac{g_1^2+ g_2^2}{2}\ \frac{\nuvev^2}{M_{\tilde Z}}
\l{neutrino-mass}
\ee
(see \cite{Hall-Suzuki,bhh}).
In the model under consideration
this contribution to tau neutrino mass is
typically larger 
than the one produced by the loop-diagram stipulated by
the interaction \refe{lambda333}. 

We can derive from \refe{neutrino-mass} the  bound on 
$R$-parity violating couplings. Taking into account that  
$\leff\sim \theta_\D\ y_b,$ and  
$\nuvev \sim v\ \theta_\D\ y_b^2$ we get the relation between 
$\leff$ and neutrino mass 
\be
\leff\sim 0.06\times 
\left[\frac{\theta_\D}{0.1\ {\rm rad.}} \right]^{1/2}
\!\!\!\times \left[\frac{m_{\nu_\tau}}{10\ {\rm MeV}} \right]^{1/4}
\!\!\!\times \left[\frac{M_{\tilde Z}}{1\ {\rm TeV}} \right]^{1/4}.
\ee
Therefore it is possible to 
obtain large $R$-parity violating couplings
with tau neutrino masses close to the 
present experimental limit.
For $m_{\nu_\tau}= {\cal O} (30$ eV), corresponding to 
the cosmological bound on stable $\nu_\tau$, 
the coupling $\leff$ becomes
of the order 0.002. 
For such values of $\leff$
the detection of supersymmetric particle decays is 
still possible: the condition to be satisfied is in fact
$\lambda_{\rm obs}'$ 
$\gtap 2\cdot 10^{-5}$  
$\sqrt{\gamma}$
$({\tilde m}/1\ {\rm TeV})^2$ 
$(150\ {\rm GeV}/ m_{\chi})^{5/2},$ 
where $\gamma$ is the Lorentz
boost factor \cite{Dreiner}.

\section{Discussion and conclusions}

The $R$-parity breaking couplings offer great 
possibilities for phenomenological speculations, 
but, up to now, no effect which should be related 
to them has been found.
This may be due to the fact that $R$-parity breaking 
couplings are small; in this case one could 
expect physical effects in rare or forbidden processes.  
But, just on the basis of the observed phenomena, 
certain $R$-parity breaking couplings 
may be large.
This unclear situation calls either for further 
theoretical or experimental informations.
It is encouraging that rather clear patterns for 
$R$-parity breaking couplings emerge
in the context of supersymmetric Grand Unification.
Models in which both lepton and baryon-violating
couplings are small have been discussed.
Large $R$-parity breaking couplings are present in
another kind of models, based on the doublet-triplet splitting.
In the context of the low-energy supergravity models
for supersymmetry breaking, we pointed to   
an interesting signature of this second scenario: 
the correlation between the size of the $R$-parity 
breaking coupling and 
the mass of the tau neutrino.
\section*{Acknowledgments}
A large part of the material exposed is based on my collaboration
with A.Yu.\ Smirnov, whom I thank most warmly.
I would like also to thank the Organizers, and in particular
N.D.\ Tracas, for the pleasant and stimulating atmosphere they were
able to create for the $5^{th}$ {\em Hellenic School and Workshops
on Elementary Particle Physics}
(I would say in accord with the spirit of \cite{ERWIN}).
Finally I take the opportunity 
to acknowledge pleasant and useful 
conversations with
M.\ Bastero-Gil, Z.\ Berezhiani, B.\ Brahmachari,
H.\ Dreiner, G.\ Dvali, P.\ Fayet, G.\ Fiorentini,
G.\ Leontaris, A.\ Melfo, N.\ Paver, E.\ Roulet, 
C.\ Savoy, G.\ Senjanovi\'c
and J.\ Steinberger.


\end{document}